\documentclass[slac,twocolumn,twoside]{revtex4}
\usepackage{fancyhdr}
\begin{document}

\title{Pitfalls of Goodness-of-Fit from Likelihood}

\author{Joel Heinrich}
\affiliation{University of Pennsylvania, Philadelphia, PA 19104, USA}

\begin{abstract}
The value of the likelihood is occasionally used by high energy
physicists as a statistic to measure goodness-of-fit in unbinned
maximum likelihood fits.  Simple examples are presented that
illustrate why this (seemingly intuitive) method fails in practice to
achieve the desired goal.
\end{abstract}

\maketitle

\thispagestyle{fancy}


\section{INTRODUCTION}
\begin{quote}
For every complex problem, there is a solution that is simple, neat, and wrong.
\textit{H.L.~Mencken}
\end{quote}

The complex problem considered here is goodness-of-fit (g.o.f.)\ for
unbinned maximum likelihood fits in cases when binned g.o.f.\ methods
and Kolmogorov-Smirnov are not well suited:

A physicist, having fit a complicated model to his multi
dimensional data to obtain estimates of the values of certain
parameters, is also expected to check how well the data match his
model.  In the sections that follow, we discuss a g.o.f.\ method,
still occasionally used in high energy physics (HEP), that is simple,
neat, and wrong.

\section{THE SNW\protect\footnotemark\ METHOD}

\footnotetext{Simple, Neat, Wrong.}
We start with a brief description of the method. (A true derivation,
for obvious reasons, is not available.)

\begin{description}
\item[observation:] Maximum likelihood fits are performed
by maximizing the likelihood $L(\vec\theta,\vec x)$ with respect to the
(unknown) parameters $\vec\theta$ for fixed data $\vec x$.
\item[faulty intuition:] Thus, the value of the likelihood provides
the g.o.f.\ between the data and the probability density function
(p.d.f.):  The value of the likelihood at the maximum,
\[L_\mathrm{max}=L(\vec{\hat\theta},\vec x)\]
corresponds to the best
fit---the smaller the likelihood, the worse the g.o.f., \ldots
\item[obstacle:] To calculate this ``g.o.f.''\ P-value, we need the
distribution of $L_\mathrm{max}$ for
an ensemble of random $\vec x$
deviates from the p.d.f.\ using the true (but unknown) parameters
$\vec\Theta$.
\item[faulty resolution:] We approximate this by
replacing $\vec\Theta$ with the parameter
estimate obtained from the fit to the actual data.
\end{description}

This method has a long history of use in high energy physics.
It's recommended by several excellent statistical data
analysis texts written by (and for) high energy particle physicists.
Consequently, and because the method is ``obvious'',
it's still being used in (some) HEP analyses.

Reference \cite{eadie}, written by a
statistician and four physicists, describes the method, but
criticizes:
\begin{quote}
The likelihood of the data would appear to be a good
[g.o.f.] candidate at
first sight.  Unfortunately, this carries little information as a test
statistic, as we shall see\ldots
\end{quote}
Since this was ignored, maybe its warning was not strong enough.
I have found no mention of the
method in texts written (solely) by statisticians.

\section{A SIMPLE TEST OF THE METHOD}
\begin{quote}
Always test your general reasoning against simple models.
\textit{John S.~Bell}
\end{quote}

Reference \cite{cdf5639}, following the above advice,
tests the method against the p.d.f.
\[
{1\over\tau}e^{-t/\tau} \qquad (t\ge0)
\]
where $t$ (we have in mind the decay-time of a particle) follows an
exponential distribution, and $\tau$ (the mean lifetime)
is a parameter whose value, being unknown, is estimated from data.
The likelihood for $N$ observations $t_i$ is
given by
\[
-\ln L=\sum_{i=1}^N\left[\ln\tau+{t_i\over\tau}\right]
\]
The value ($\hat\tau$) of $\tau$ that maximizes the likelihood, and the
value ($L_\mathrm{max}$) of the likelihood at its maximum, are given by
\[
{\hat\tau={1\over N}\sum_{i=1}^Nt_i} \qquad
{-\ln L_\mathrm{max}=N(1+\ln\hat\tau)}
\]

\subsection{The First Surprise\label{s1}}
The value of the likelihood at its maximum (in this test case) is just
a simple function of $\hat\tau$---all samples with the same mean
obtain the same ``g.o.f.''\ value. This is a disaster for g.o.f. Even
if the true value of $\tau$---call it $\mathcal{T}$---were known in
advance, so that we could calculate the P-value associated with the
observed $\hat\tau$, merely comparing the $\hat\tau$ of
the data with $\mathcal{T}$ is not sufficient to
show that the observed data are modeled well by the exponential
distribution.

\subsection{The Second Surprise}
Since under this method, our P-value ensemble is actually based on the
value of $\hat\tau$ computed from the data (not knowing the true value
$\mathcal{T}$), we \emph{always} obtain a P-value of about 50\%,
\emph{for any data whatsoever}. This is a second disaster for g.o.f.
By construction, the distribution of $L_\mathrm{max}$ from our
ensemble of $N$-event pseudo experiments tracks the $L_\mathrm{max}$
observed from the data.

The fact that the method yields ``reasonable'' P-values has
undoubtedly contributed to its longevity in practice: P-values very
near 0 or 100\% would have triggered further investigation.

\subsection{Lessons Learned}
In this example, g.o.f.\ is equivalent to testing the single
hypothesis: ``The data are from an exponential distribution of
unspecified mean.'' $L_\mathrm{max}$ provided no information 
with respect to this hypothesis.

What went wrong?  In our test case, the likelihood could be expressed
as a function of just the parameter and its maximum likelihood
estimator (m.l.e.): $L(\tau;\hat\tau)$.  \emph{All} data samples with
the same m.l.e.\ gave the same ``g.o.f.''

Exactly the same thing happens in the Gauss\-ian (normal) case---the
likelihood can be written using solely the 2 parameters and their
estimators: $L(\mu,\sigma;\hat\mu,\hat\sigma)$.

Other ``textbook'' distributions---scaled gamma, beta,
log-normal, geometric---also fail in the same way.
Geometric is a discrete distribution, so the problem is not
restricted to the continuous case.

\section{MORE TROUBLE: NON INVARIANCE}
Returning to our exponential example, suppose we make the substitution
$t=x^2$. The p.d.f.\ transforms as
\[{1\over\tau}e^{-t/\tau}dt={2x\over\tau}e^{-x^2/\tau}dx\]
and the g.o.f.\ statistic is now calculated as
\[-\ln L=\sum_{i=1}^N\left[\ln\tau+{x_i^2\over\tau}-\ln(2x_i)\right]
\qquad
{\hat\tau={1\over N}\sum_{i=1}^Nx_i^2}
\]
\begin{eqnarray*}
-\ln L_\mathrm{max}&=&
{N(1+\ln\hat\tau)-\sum_{i=1}^N\ln(2x_i)}\\
&=&{N(1+\ln\hat\tau)-\sum_{i=1}^N\ln(2\sqrt{t_i})}
\end{eqnarray*}
That is, the ``g.o.f.''\ statistic is not invariant under change of
variable in the continuous p.d.f.\ case. (The value of the m.l.e.\ is,
of course, invariant.)

Under change of variable, the ``g.o.f.''\ statistic picks up an extra
term from the Jacobian---an extra function of the data.  We're free to
choose any transformation, so we can make the ``g.o.f.''\ statistic
more or less anything at all---a serious pathology.

At this point, experts point out that \emph{ratios} of likelihoods
have the desired invariance under change of variable, but, while the
likelihood ratio is a useful test statistic in certain special cases,
it is not at all clear how to obtain a useful g.o.f.\ statistic from
the likelihood ratio in the general, unbinned, case.

\section{A REPLACEMENT MODEL}
Since we now lack an intuitive understanding, we need a replacement
intuition for what is going on. I propose this model:

Denote by $H_0$ the hypothesis that the data are from the p.d.f.\ in
question.  Specify an alternative hypothesis $H_1$ that the data are
from a uniform p.d.f.\ (flat in the variables that we happen to have
chosen).  At least, the $H_1$ p.d.f.\ is flat over the region where we
have data---outside that region it can be cut off.

Performing a classic Neyman-Pearson hypothesis test of $H_0$ vs $H_1$,
we use the ratio  of their likelihoods as our test statistic:
\[\lambda(\vec x)=
{L(\vec x|H_0)\over L(\vec x|H_1)}={L(\vec x)\over\mathrm{constant}}\]
So, the  ``g.o.f.''\ statistic can be re-interpreted as suitable for
a hypothesis test that indicates which of $H_0$ (our p.d.f.)\ and $H_1$
(a flat p.d.f.)\ is more favored by the data---a well established
statistical practice.

The benefit of the new interpretation is that it explains behaviors
that were baffling under the g.o.f.\ interpretation: Neyman-Pearson
hypothesis tests and g.o.f.\ tests behave quite differently.

For example, a reasonable g.o.f.\ statistic should be at least
approximately distribution independent, but $\lambda(\vec x)$ is often
highly correlated with the m.l.e.'s (100\% in our exponential
case). This high correlation was confirmed in the example contributed
by K.~Kinoshita\cite{kinoshita} to the 2002 Durham Conference.  Not
knowing the true value of the parameters then makes it difficult, or
impossible, to use $\lambda(\vec x)$ as g.o.f., since we don't know
what $\lambda(\vec x)$
\emph{should} be.\footnote{Small correlations are not fatal.  For example, if
the \mbox{P-value} of g.o.f.\ for the observed data in a particular
case ranged only between, say, 20\% and 30\%, for different true values
within $\pm3\sigma$ of the estimated value of a parameter, one would
be justified in concluding ``good fit'' (assuming the g.o.f.\
statistic used had the right properties in other respects).} The
behavior of these correlations is natural and obvious in the
hypothesis test picture: changing the parameters changes the
``flatness'' of the $H_0$ p.d.f., and $\lambda(\vec x)$ reflects this.

Reference \cite{eadie} pointed out that, with no unknown parameters,
one can always transform the p.d.f.\ to a flat distribution.  Then
$\lambda(\vec x)$ becomes constant independent of the data---bad news
for g.o.f.  In the hypothesis test picture, this becomes a comparison
between two identical hypotheses, and the result is what we would
expect.

\section{TEST BIAS\label{bias}}
Take the $H_0$ p.d.f.\ to be
\[
e^{-t} \qquad (t\ge0)
\]
This distribution is fully specified---no unknown parameters.
Our ``g.o.f.''\ statistic is then
\[
-\ln L=N\hat t
\]
whose mean is
$\langle-\ln L\rangle=N$, and variance is
$\mathrm{Var}(-\ln L)=N$, for an ensemble of data sets
from the $H_0$ p.d.f. A data set with $\hat t$ close enough to 1 will
be claimed to be a good fit to the $H_0$ p.d.f.

But say, unknown to us, the data are really from a triangular
p.d.f.:
\[
1-|t-1| \qquad (0\le t\le2)
\]
The mean and variance of $N\hat t$
will be $N$ and $N/6$ respectively, for data from the triangular
distribution. So,
although the exponential and triangular p.d.f.'s are quite different,
the triangular data will be more likely to pass the g.o.f.\ test than
exponential data for which it was intended.
Statisticians refer to this situation as a case of ``test bias''.

 We conclude that, even with no free parameters, the
``g.o.f.''\ test is biased: there exist ``impostor'' p.d.f.'s that
should produce bad fits, but instead pass the ``g.o.f.''\ test with greater
probability then the p.d.f.\ for which the test was designed.
Reference \cite{cdf6123} gives additional examples of this behavior.

From the hypothesis test point of view, this behavior makes sense.
The exponential and triangular data have the same ``distance'' from
the flat distribution, on the average, with the triangular data being
less susceptible to fluctuations.  The hypothesis test doesn't tell us
when the data are inconsistent with both $H_0$ and $H_1$.

\section{ANOTHER EXAMPLE}
Here we try to find an example p.d.f.\ (with a free parameter) that
the method in question can handle well. We use the insight provided by the
hypothesis test picture. We want to keep the correlation between the
free parameter and the g.o.f.\ statistic $L_\mathrm{max}$ to a
minimum.  In the hypothesis test picture, this is achieved when the
``flatness'' of the p.d.f.\ is independent of the parameter.  A
location parameter has this property.  Additionally, we want the
p.d.f.\ to be easily distinguishable from a flat p.d.f.
So we choose the Gaussian
\[
{1\over\sqrt{2\pi}\sigma}e^{-0.5(x-\mu)^2/\sigma^2}
\]
where $\mu$ is unknown, but $\sigma$ is specified in advance.
The likelihood is given by
\[
-\ln L=\sum_{i=1}^N\left[\ln\sqrt{2\pi}+\ln\sigma+
{1\over2}\left({x_i-\mu\over\sigma}\right)^{\!\!2}\right]
\]
When $\mu$ and $\sigma$ are both unknown, their m.l.e.'s are
\[
\hat\mu={1\over N}\sum_{i=1}^Nx_i \qquad
\hat\sigma^2={1\over N}\sum_{i=1}^N(x_i-\hat\mu)^2
\]

Using these expressions, we can rewrite the likelihood in the form
$L(\mu,\sigma;\hat\mu,\hat\sigma)$:
\[
-\ln L=
{N\over2}\left[\ln(2\pi)+\ln(\sigma^2)+
{\hat\sigma^2+(\hat\mu-\mu)^2\over\sigma^2}\right]
\]
When only $\mu$ is unspecified, its m.l.e.\ is $\hat\mu$ as above, and
the value of the maximized likelihood is
\[
-\ln L_\mathrm{max}=
{N\over2}\left[\ln(2\pi)+\ln(\sigma^2)+{\hat\sigma^2\over\sigma^2}\right]
\]

Our victory is that $L_\mathrm{max}$ only depends on
$\hat\sigma$, which is an ancillary statistic for $\mu$. That is,
we don't need to know the true value of $\mu$ in order to calculate
the distribution of our g.o.f.\ statistic in this carefully
chosen example. In fact, a convenient form for the g.o.f.\ statistic is
\[
N{\hat\sigma^2\over\sigma^2}=
\sum_{i=1}^N\left({x_i-\hat\mu\over\sigma}\right)^{\!\!2}
\]
which is well known to have the distribution (under the null
hypothesis) of a $\chi^2$ with $N-1$ degrees of freedom.

\subsection{The Bad News}
Before we declare that the method performs well in this example,
there are several ugly facts to consider:
\begin{itemize}

\item Data that match the null hypothesis well yield
$N\hat\sigma^2/\sigma^2\simeq N$.  Much larger or much smaller values
of the g.o.f.\ statistic imply poor g.o.f. This is in contrast to
Pearson's $\chi^2$ (binned $\chi^2$), for example, where smaller
$\chi^2$ is always better g.o.f. So we must interpret this statistic
differently than how we are used to.

\item The g.o.f.\ in this example simply reduces to a comparison
between the sample variance and $\sigma^2$.  Any distribution with
variance approximately equal to $\sigma^2$ will usually generate data
that ``pass the test'', even distributions that look nothing like a
Gaussian. This is the same kind of problem that we first saw in
section~\ref{s1}.

\item A construction similar to that of section~\ref{bias}
will produce ``impostor'' p.d.f.'s that pass the ``g.o.f.''\ test
with greater frequency than the null hypothesis. So, we have
not eliminated the test bias problem.

\end{itemize}

In this example, the g.o.f.\ method in question will be able to flag
some, but not all, of data samples that poorly match the null
hypothesis. In answer to the question ``Are the data from a Gaussian
with unspecified mean, and variance equal to $\sigma^2$?'', this
g.o.f.\ method can only answer ``No'' or ``Maybe'': it checks the
variance part of the question, but does nothing to check the Gaussian
part.

\section{CONCLUSIONS}
\begin{itemize}
\item This ``g.o.f.''\ method is fatally flawed in the unbinned case.
Don't use it. Complain when you see it used.

\item With fixed p.d.f.'s, the method suffers from test bias, and
is not invariant with respect to change of variables. These
problems persist when there are floating parameters.

\item With floating parameters, the method is often circular:
``g.o.f.''\ becomes a comparison between the measured values
and the true (but unknown) values of the parameters\ldots

\item The misbehavior of this ``g.o.f.''\ statistic is understandable when
reinterpreted as the ratio between the likelihood in question and a
uniform likelihood, and used to distinguish between these two specific
hypotheses. Dual-hypothesis tests are not g.o.f.\ tests.
\end{itemize}

\begin{acknowledgments}
I would like to thank Louis Lyons for several helpful
discussions of the points raised here, and the organizers of the
PHYSTAT2003 Conference for arranging a superb program.
\end{acknowledgments}


\end{document}